\documentclass[amsmath,amssymb,nofootinbib]{revtex4}
\usepackage{graphicx}
\usepackage{dcolumn}
\usepackage{color}
\usepackage{bm}
\usepackage{amsmath,amssymb,graphicx}
\usepackage{epsfig}
\usepackage{enumerate}
\usepackage{array}
\usepackage{multirow}

\newtheorem{theorem}{Theorem}

\newtheorem{prop}{Proposition}

\begin{document}
\title
{Dynamical symmetries of supersymmetric oscillators}
\author{Akash Sinha\footnote{E-mail: s23ph09005@iitbbs.ac.in}$^1$,  Aritra Ghosh\footnote{E-mail: ag34@iitbbs.ac.in}$^1$, Bijan Bagchi\footnote{E-mail: bbagchi123@gmail.com}$^2$}

\vspace{2mm}

\affiliation{$^{1}$School of Basic Sciences, Indian Institute of Technology Bhubaneswar, Jatni, Khurda, Odisha 752050, India\\
$^{2}$Brainware University,
Barasat, Kolkata, West Bengal 700125, India}

\vskip-2.8cm
\date{\today}
\vskip-0.9cm


\begin{abstract}
In this paper, we describe the dynamical symmetries of classical supersymmetric oscillators in one and two spatial (bosonic) dimensions. Our main ingredient is a generalized Poisson bracket which is defined as a suitable classical counterpart to commutators and anticommutators. In one dimension, i.e., in the presence of one bosonic and one fermionic coordinate, the Hamiltonian admits a \(U(1,1)\) symmetry for which we explicitly compute the first integrals. It is found that suitable forms of the supercharges emerge in a natural way as fermionic conserved quantities. Following this, we describe classical supercharge operators based on the generalized Poisson bracket and subsequently define supersymmetry transformations. We perform a straightforward generalization to two spatial dimensions where the Hamiltonian has an overall \(U(2,2)\) symmetry. We comment on plausible supersymmetric generalizations of the Pais-Uhlenbeck and isotonic oscillators, and also present the possibility of defining a generalized Nambu bracket within the classical formalism.
\end{abstract}

\maketitle

\section{Introduction}
Supersymmetric quantum mechanics (SQM) has emerged as an area of active research (see for example, Refs. \cite{rau, junker, SUSYRev, SUSYBook, miel2004, fer2004, dong2007, fer2010, and2012, MZ}) since the seminal papers by Witten \cite{Witten1,Witten2}. The formalism of SQM rests on the supplementation of fermionic coordinates \cite{SUSYRev,SUSYBook} to the usual bosonic ones with certain graded transformations operating between them. This leads to the emergence of supercharges that act as the generators of the supersymmetry (SUSY) transformations and therefore in a sense, characterize the whole theory. The supercharges commute with the underlying Hamiltonian defining SQM and are described by a `super' algebra. For a system with one bosonic and one fermionic degree of freedom, one has the supercharge (fermionic themselves) operators \(Q\) and \(\overline{Q}\), whose anticommutator defines the Hamiltonian of the system, i.e.,
\begin{equation}\label{a}
\left[ Q , \overline{Q} \right]_+ = 2\hbar H,
\end{equation}
while \([Q,H]_- = [\overline{Q},H]_- = 0\). Here, $[\cdot,\cdot]_{-/+}$ denotes the commutator and the anticommutator, respectively. Although one usually sets $\hbar=1$, here we retain it explicitly as the presence of \(\hbar\) in the underlying equations facilitates ascertaining their corresponding behavior in the limit $\hbar \rightarrow 0$. One could now ask a pertinent question as to what would be the appropriate classical limit of a SQM theory. Since the quantum theory involves both commutators and anticommutators, the corresponding classical theory must incorporate a general bilinear-bracket structure in the space of observables which could be realized as the \(\hbar \rightarrow 0\) limit of the quantum theory. Such a bracket was introduced in \cite{soni} (see also \cite{SUSYBook}), and has been termed as a `generalized' Poisson bracket. It relates to the corresponding quantum version as
\begin{eqnarray}
    \lim_{\hbar\to 0}\frac{\left[{\cal F},{\cal T}\right]}{i\hbar}=\{{\cal F},{\cal T}\},
\end{eqnarray}
where $\{\cdot,\cdot\}$ denotes the `generalized' Poisson bracket and $\left[{\cal F},{\cal T}\right]={\cal F}{\cal T}-(-1)^{\Pi({\cal F})\Pi({\cal T})}{\cal T}{\cal F}$. The quantity $\Pi({\cal F})$ is called the parity of ${\cal F}$ and is defined in Sec. (\ref{sec2}). It is noteworthy that the corresponding `classical' theory  incorporates both anticommuting as well as c-number variables in the form of coordinates
and momenta establishing a mathematical framework dubbed as pseudomechanics in \cite{pseudo}. \\

The necessity of using the anticommutating variables arises from the requirement of preserving the relevant algebraic structure when taking the limit $\hbar\to 0$. Starting from the quantum-field-theoretic description where the fermionic fields are quantized using canonical anticommutation relations (CAR), one requires, in the limit $\hbar\to 0$, the fermionic fields to be described by anticommutating Grassmann variables in order to preserve the algebraic relations, including the equations of motion. Thus, to describe classical dynamics on the fermionic phase space, it only seems reasonable to describe the phase space coordinates themselves using Grassmann numbers. A particular state of the system is then given by a point on the phase space and the observables are the functions of the phase space variables. As will be discussed later in some detail, the notion of fermionic variables leads to the concepts of left and right derivatives \cite{soni}. \\

The present work is related to an earlier paper \cite{aniso} by the present authors. The motivation for this work is to expose the symmetry structure of supersymmetric oscillators within a classical framework. In particular, we focus on the scenario of pseudo-classical mechanics or pseudomechanics of supersymmetric oscillators with respect to the notion of the generalized Poisson bracket. Towards this end, we formulate a pseudomechanical framework, closely related to the first-order formalism of SQM wherein all the bilinear-bracket relations are described in terms of the generalized Poisson bracket. For harmonic oscillators in one and two spatial (bosonic) dimensions, we show that the supersymmetric Hamiltonian has an underlying \(U(1,1)\) and \(U(2,2)\) symmetry, respectively, such that the supercharges that describe the SUSY transformations between the bosonic and fermionic sectors emerge as conserved quantities. Furthermore, we propose supersymmetric generalizations of a certain positive-energy case of the Pais-Uhlenbeck oscillator and also the isotonic oscillator, both of which are well known for their appearance in physical problems. 

\section{Generalized Poisson brackets and Hamiltonian formalism}\label{sec2}
In this section, we briefly review the basic tenets of generalized Poisson brackets \cite{SUSYBook,soni}. Consider a classical system with phase-space variables \((q,p)\). To this we append a fermionic coordinate \(\theta\) possessing an associated momentum \(\pi\). We aim to construct a classical model in which the following Poisson-bracket relations are embedded:
\begin{equation}
\{q,p\} = 1, \hspace{7mm} \{\theta,\pi\} = 1,
\end{equation} while others vanish. To this end, let us define the notions of even and odd operators which appear in the quantum theory. For a generic operator \(A\), the permutation operator \(\mathfrak{P}\) is such that \(\mathfrak{P}^{-1} A \mathfrak{P} = (-1)^{\Pi(A)} A\), where \(\Pi(A) = 0\) if \(A\) is even while \(\Pi(A) = 1\) if \(A\) is odd. Since fermionic variables are `odd' with respect to permutations, it is important to consider even and odd functions with care. We preserve the same definition of odd-ness and even-ness in the classical theory, in that all the bosonic variables are even, while the fermionic ones are odd. Quite naturally then, the products of two variables, i.e., \(q^2\), \(p^2\), and \(qp\) are even. Similarly \(\theta^2(=0)\), \(\pi^2(=0)\), and \(\theta \pi\) are even too. The Hamiltonian function is defined via the generalized Legendre transform that reads as
\begin{equation}\label{Ham1}
H = -L +  \pi  \dot{\theta}+ p \dot{q} . 
\end{equation} 
Since the Lagrangian is even, while \((q,p)\) and \((\theta,\pi)\) are pairwise even and odd, respectively, clearly \(H\) is even in character.

\subsection{Generalized Poisson brackets}
For any function \(F(\mathcal{Q},\mathcal{P})\) existing on the phase space, with \(\mathcal{Q}\) and \(\mathcal{P}\) collectively denoting the coordinates and momenta respectively (both for the bosonic and fermionic variables), we define 
\begin{equation}
dF = F_{,\mathcal{Q}} d\mathcal{Q} + d\mathcal{P} \partial_\mathcal{P} F,
\end{equation} where \(F_{,\mathcal{Q}}\) stands for the right derivative while \( \partial_\mathcal{P} F\) corresponds to the left derivative. In what follows we shall adhere to the convention of treating a derivative with respect to the coordinates as the right derivative while viewing a derivative with respect to the momenta as the left derivative. Of course, the right derivatives can be permuted to the left position. Explicitly, we express 
\begin{equation}
\partial_\mathcal{Q} F  = (-1)^{\Pi(\mathcal{Q})[\Pi(\mathcal{Q}) + \Pi(F)]} F_{,\mathcal{Q}}.
\end{equation}
Now, for a Hamiltonian system, there exists a Hamiltonian vector field \(X_H(\cdot)\), such that \(X_H(F) = \dot{F}\). 
Varying the Hamiltonian in Eq. (\ref{Ham1}) results in the following equations of motion:
\begin{equation}
\dot{q} = \frac{\partial H}{\partial p}, \quad \dot{p} = -\frac{\partial H}{\partial q}, \quad \dot{\theta} = \frac{\partial H}{\partial \pi}, \quad \dot{\pi} = \frac{\partial H}{\partial \theta},
\end{equation} where we have exploited the fact that the Hamiltonian is an even function. In fact, this leads to the following expression for the Hamiltonian vector field:
\begin{equation}\label{XH}
X_H(F) = \{F, H\} = F_{,\mathcal{Q}} \partial_\mathcal{P} H - H_{,\mathcal{Q}} \partial_\mathcal{P} F,
\end{equation} or \(X_H(\cdot) = \{\cdot, H\}\). Here, \(\{\cdot,\cdot\}\) is called the generalized Poisson bracket which was introduced by employing the properties of left and right differentiation. As has been argued in \cite{soni}, this can be realized as the classical limit of SQM theory and takes into account the algebraic properties analogous to both commutator and anticommutator brackets. Moreover, this bracket satisfies a generalized Jacobi identity (see \cite{soni} for other algebraic properties):
\begin{eqnarray}
\{f,\{g,h\}\} &+& (-1)^{\Pi(f)(\Pi(g) +\Pi(h))} \{g,\{h,f\}\}  + (-1)^{\Pi(h)(\Pi(f) +\Pi(g))} \{h,\{f,g\}\} = 0, \label{jacobi} 
\end{eqnarray} for any three functions \(f\), \(g\), and \(h\) on the phase space. \\

We shall denote a generic even function by \(E\), i.e., \(\Pi(E) = 0\), while an odd function shall be denoted by \(O\), i.e., \(\Pi(O) = 1\). Subscripts shall be used to distinguish between multiple even/odd functions in the same equation. Although the expression for the generalized Poisson bracket indicated in Eq. (\ref{XH}) involves the fact that the Hamiltonian is an even function, one may write an analogous expression for the bracket between two odd functions. The general expressions are summarized as

\begin{eqnarray}
\{E_1,E_2\}_{q,p,\theta,\pi}&=&\left(\frac{\partial E_1}{\partial q}\frac{\partial E_2}{\partial p}-\frac{\partial E_2}{\partial q}\frac{\partial E_1}{\partial p}\right) +\left(-\frac{\partial E_1}{\partial \theta}\frac{\partial E_2}{\partial \pi}+\frac{\partial E_2}{\partial \theta}\frac{\partial E_1}{\partial \pi}\right), \label{EE} \\ 
\{E,O\}_{q,p,\theta,\pi}&=&\left(\frac{\partial E}{\partial q}\frac{\partial O}{\partial p}-\frac{\partial O}{\partial q}\frac{\partial E}{\partial p}\right) -\left(\frac{\partial E}{\partial \theta}\frac{\partial O}{\partial \pi}+\frac{\partial O}{\partial \theta}\frac{\partial E}{\partial \pi}\right), \label{EO} \\ 
\{O,E\}_{q,p,\theta,\pi}&=&\left(\frac{\partial O}{\partial q}\frac{\partial E}{\partial p}-\frac{\partial E}{\partial q}\frac{\partial O}{\partial p}\right) +\left(\frac{\partial O}{\partial \theta}\frac{\partial E}{\partial \pi}+\frac{\partial E}{\partial \theta}\frac{\partial O}{\partial \pi}\right), \label{OE}\\ 
\{O_1,O_2\}_{q,p,\theta,\pi}&=&\left(\frac{\partial O_1}{\partial q}\frac{\partial O_2}{\partial p}+\frac{\partial O_2}{\partial q}\frac{\partial O_1}{\partial p}\right) +\left(\frac{\partial O_1}{\partial \theta}\frac{\partial O_2}{\partial \pi}+\frac{\partial O_2}{\partial \theta}\frac{\partial O_1}{\partial \pi}\right). \label{OO}
\end{eqnarray}

\subsection{Structure of oscillator Hamiltonians} 
Let us examine the one-dimensional bosonic oscillator:
\begin{equation}
H_{\rm B} = \frac{p^2 + q^2}{2},
\end{equation} where \((q,p)\) are a set of real variables, satisfying \(\{q,q\} = \{p,p\} = 0, \{q,p\} =1\). We define a pair of complex-valued variables \((X,P)\) according to \cite{aniso,ComplexCan} 
\begin{equation}\label{can_pq}
X = \frac{q - ip}{\sqrt{2}}, \hspace{5mm} P = \frac{p - iq}{\sqrt{2}}.
\end{equation}
 Using the coordinate expressions for the generalized Poisson bracket as given by Eqs. (\ref{EE})-(\ref{OO}), it is easy to verify the canonical nature of the transformation, namely,
\begin{eqnarray}
\{X,X\}_{q,p}=0=\{P,P\}_{q,p},\quad\quad \{X,P\}_{q,p}=1.
\end{eqnarray} The transformations [Eq. (\ref{can_pq})] can be interpreted as the classical analogue of Bogoliubov transformations; notice that \(P = - i X^*\). In terms of the new variables as given by Eq. (\ref{can_pq}), the Hamiltonian turns out to be 
\begin{equation}\label{Hbos}
H_{\rm B} = i PX,
\end{equation} which is a modified form of the harmonic oscillator in terms of the $(X, P)$ coordinates. Note that because of the aforementioned condition \(P = - i X^*\), the number of independent variables is effectively reduced to a pair of real ones. Similarly, the fermionic oscillator is guided by the Hamiltonian \cite{soni} 
\begin{equation}\label{Hfer}
H_{\rm F} = i \pi \theta,
\end{equation} with the requirement that \(\pi = i \theta^*\) where both \(\theta\) and \(\pi\) are Grassmann numbers. 

\section{One-dimensional supersymmetric oscillator}\label{sec3}
The Hamiltonian of a classical supersymmetric oscillator is described by the simple superposition, \(H = H_{\rm B} + H_{\rm F}\), where \(H_{\rm B}\) is given by (\ref{Hbos}), while \(H_{\rm F}\) is defined by Eq. (\ref{Hfer}). Thus, 
\begin{eqnarray}\label{Ham1d}
H=i (PX + \pi \theta),
\end{eqnarray} 
and the accompanying phase space \(\mathcal{M}\) can be split up as \(\mathcal{M} = \mathcal{M}_B \oplus \mathcal{M}_F\), where \(\mathcal{M}_B\) and \(\mathcal{M}_F\) contain the variables \((X,P)\) and \((\theta,\pi)\), respectively.\\

To build up a supersymmetric structure, let us introduce two fermionic quantities, $Q$ and $\overline{Q}$, which are odd in character and are defined as

\begin{eqnarray}
Q=\alpha P\theta +\beta X\pi, \quad\quad\quad \overline{Q}=\epsilon P\theta +\delta X\pi,
\end{eqnarray}
where $\{\alpha,\beta,\epsilon,\delta\}$ are arbitrary nonzero complex constants. We restrict $\frac{\alpha}{\epsilon}\neq\frac{\beta}{\delta}$ to avoid $Q$ and $\overline{Q}$ becoming linearly dependent. Imposing $\overline{Q}=Q^*$, we find \begin{eqnarray}
Q=\alpha P\theta +\beta X\pi, \quad\quad\quad \overline{Q}=\left(\beta ^* P\theta +\alpha ^* X\pi\right),
\end{eqnarray}
and the condition for the linear independence of \(Q\) and \(\overline{Q}\) points to $|\alpha|^2\neq|\beta|^2$. It is easy to check $\{Q,\overline{Q}\}=(\alpha \delta+\beta \epsilon)(PX+\pi\theta)$, and thus, for $\overline{Q}=Q^*$, we have $(\alpha \delta+\beta \epsilon)=(|\alpha|^2+|\beta|^2)\neq 0$. This gives
\begin{eqnarray}\label{QQH11}
\{Q,\overline{Q}\}\sim H,
\end{eqnarray} analogous to Eq. (\ref{a}) where \(\{\cdot,\cdot\}\) denotes the generalized Poisson bracket as opposed to the anticommutator that appears in Eq. (\ref{a}).
We interpret \(Q\) and \(\overline{Q}\) as the candidates for the `classical supercharges'. These are also subject to the following nilpotent relations:
\begin{eqnarray}\label{2}
\{Q,Q\}=0=\{\overline{Q},\overline{Q}\}.
\end{eqnarray}
The above conditions are fulfilled for certain typical choices of the parameters $(\alpha,\beta)$. For instance, if one takes $\alpha=\sqrt{2}$ and $\beta=0$, then the conditions in Eq. (\ref{2}) are obeyed and further,
\begin{eqnarray}\label{QQH}
\frac{i}{2}\{Q,\overline{Q}\}=H,
\end{eqnarray} which is the pseudomechanical analogue of Eq. (\ref{a}). As can be verified easily, one has
\begin{eqnarray}
\{Q,\overline{Q}\}=\frac{2 H}{i}=\lim_{\hbar\to 0}\frac{2 \hbar H}{i\hbar}=\lim_{\hbar\to 0}\frac{\left[Q,\overline{Q}\right]}{i\hbar},
\end{eqnarray}
as expected. The choice of the parameters is by no means unique; Eq. (\ref{QQH}) is also recovered for \(\alpha = 0\) and \(\beta = \sqrt{2}\). \\

Now, with any choice of \(\alpha \neq 0\) and \(\beta = 0\), one can verify that
\begin{eqnarray}
&&\{X,Q\}\sim\theta,\quad \{P,Q\}=0,\quad\{\theta,Q\} = 0,\quad \{\pi,Q\}\sim P, \label{17} \\
&&\{X,\overline{Q}\} = 0,\quad \{P,\overline{Q}\}\sim\pi,\hspace{3.2mm}\{\theta,\overline{Q}\}\sim X,\hspace{2.3mm} \{\pi,\overline{Q}\} = 0, \nonumber 
\end{eqnarray}
where all the generalized Poisson brackets are evaluated in the $(X,P,\theta,\pi)$ basis and the symbol `\(\sim\)' indicates that the relations are true up to a constant factor, proportional to \(\alpha\). Eq. (\ref{17}) is a set of SUSY transformations, generated by \(Q\) and \(\overline{Q}\) for \(\alpha \neq 0\) and \(\beta = 0\). Similarly, if one chose \(\alpha = 0\) and \(\beta \neq 0\), then Eqs. (\ref{QQH11}) and (\ref{2}) are still satisfied but Eq. (\ref{17}) would become
\begin{eqnarray}
&&\{X,Q\} = 0,\quad \{P,Q\}\sim \pi,\quad\{\theta,Q\} \sim X,\quad \{\pi,Q\} = 0, \label{18} \\
&&\{X,\overline{Q}\} \sim \theta,\quad \{P,\overline{Q}\}=0,\quad\hspace{0.4mm}\{\theta,\overline{Q}\} = 0,\quad\hspace{1.2mm} \{\pi,\overline{Q}\} \sim P, \nonumber 
\end{eqnarray} where the symbol `\(\sim\)' indicates that the relations are true up to a constant factor, proportional to \(\beta\). Eq. (\ref{18}) is the set of SUSY transformations for \(\alpha = 0\) and \(\beta \neq 0\). Therefore, we are in a position to make the following proposition: 

\begin{prop}\label{prop1}
The operators \(\mathfrak{X}_Q (\cdot) = \{\cdot,Q\}\) and \(\mathfrak{X}_{\overline{Q}} (\cdot) = \{\cdot,\overline{Q}\}\) acting on the algebra of functions on the phase space define the SUSY transformations.
\end{prop}

We would interpret \(\mathfrak{X}_Q (\cdot) = \{\cdot,Q\}\) and \(\mathfrak{X}_{\overline{Q}} (\cdot) = \{\cdot,\overline{Q}\}\) as `classical supercharge operators' in the present context. The following result also holds: 

\begin{theorem}
The classical supercharge operators satisfy \(\mathfrak{X}_Q^2 = \mathfrak{X}_{\overline{Q}}^2 = 0\). 
\end{theorem}

\textit{Proof -} By a straightforward computation using the generalized Jacobi identity (\ref{jacobi}).

\subsection{Conserved quantities}
We shall now show that one actually need not define the supercharges by hand; rather they emerge as conserved quantities of the theory. Consider the following sets of quantities: $\mathcal{P}=(P\;\;\pi\xi^*)^T$ and $\mathcal{Q}=(X\;\;\xi\theta)^T$, with $\xi$ being a Grassmann number such that $\xi^*\xi=1$. One should note that the quantities $\mathcal{P}$ and $\mathcal{Q}$ are completely bosonic in nature. Then, the total Hamiltonian reads
\begin{eqnarray}\label{H1d}
H=i\mathcal{P}^T\mathcal{Q}.
\end{eqnarray}
It is not hard to see that $\mathcal{P}=-i\sigma_3 \mathcal{Q}^*$. Therefore, the Hamiltonian shall admit a $U(1,1)$ symmetry, i.e., $ \mathcal{Q}\rightarrow u \mathcal{Q},\;\mathcal{P}\rightarrow (\sigma_3 u^{*}\sigma_3)\mathcal{P}$. The conserved quantities can be calculated by demanding that under the action of $u\in U(1,1)$ \cite{u11}, the Hamiltonian remains invariant. Now, the infinitesimal variation of $\mathcal{Q}$ and $\mathcal{P}$ under the action of $u=\exp{[i\phi_{\mu} T^{\mu}]}$ is given by
\begin{eqnarray}
\delta \mathcal{Q}=i\phi_{\mu} T^{\mu} \mathcal{Q},\quad \quad \quad \delta \mathcal{P}=-i\phi_{\mu} (\sigma_3 (T^{\mu})^*\sigma_3) \mathcal{P},
\end{eqnarray} where \(\phi^{\mu}\in \mathbb{R}\) (for each \(\mu\)) and \(\mu = 0,1,2,3\). Here, the $T^{\mu}$'s are the generators of $U(1,1)$ group and are given by $T^{0}=\frac{1}{2}\sigma^0,T^{3}=\frac{1}{2}\sigma^{3}, T^{1}=\frac{i}{2}\sigma^{1}, T^{2}=\frac{i}{2}\sigma^{2}$; where $\sigma^0=I$ and $\sigma^{i},\;i=1,2,3$ are the Pauli matrices. Demanding that the infinitesimal variation of the Hamiltonian vanishes under this transformation, we get the conserved quantities to be 
\begin{eqnarray}
\mathcal{Z}^{\mu}\sim(\sigma^{\mu})_{jk}\mathcal{P}_j\mathcal{Q}_k,\quad\quad\quad \mu=0,1,2,3 .
\end{eqnarray}
It may be verified that all the conserved quantities are not independent of each other and further, we have
\begin{eqnarray}
&&\text{Bosonic Hamiltonian}\sim\frac{\mathcal{Z}^0+\mathcal{Z}^3}{2},\\ \nonumber
&& \text{Fermionic Hamiltonian}\sim\frac{\mathcal{Z}^0-\mathcal{Z}^3}{2},\\ \nonumber
&&\mathcal{Z}^1\sim(\xi Q-\xi^*\overline{Q}),\quad\quad \mathcal{Z}^2\sim - i(\xi^*\overline{Q}+\xi Q).
\end{eqnarray}
We note that all the conserved quantities are bosonic in nature. However, the fermionic conserved quantities (the supercharges) can be recovered from here as
\begin{eqnarray}
    Q\sim\frac{\xi^*}{2}\left({\cal Z}^1+i{\cal Z}^2\right),\quad \quad ~\overline{Q}\sim\frac{\xi}{2}\left({\cal Z}^1-i{\cal Z}^2\right).
\end{eqnarray}

\section{Generalization to two dimensions}\label{sec4}

\subsection{Conserved quantities}
Let us start with the following definitions:
\begin{eqnarray}
\mathcal{P}=(P_1\;\;P_2\;\;\pi_1\xi^*_1\;\;\pi_2\xi^*_2)^T,\quad \mathcal{Q}=(X_1\;\;X_2\;\;\xi_1\theta_1\;\;\xi_2\theta_2)^T, \nonumber \\
\end{eqnarray}
with $\xi^*_i\xi_i=1\text{ for } i=1,2$. The Hamiltonian can be written as 
\begin{eqnarray}\label{H2dSusy}
H=i\mathcal{P}^T\mathcal{Q}.
\end{eqnarray}
 Following the discussion of the previous section, one can see that this Hamiltonian enjoys a $U(2,2)$ symmetry. Let us denote the generators of this group by 
\begin{eqnarray}
\lambda^{\mu};\quad \mu=0,1,\cdots,15.
\end{eqnarray}
Then the conserved quantities are (a summation over the repeated indices \(j\) and \(k\) is implied)
\begin{eqnarray}\label{C2D}
\mathcal{C}^{\mu}\sim (\lambda^{\mu})_{jk}\mathcal{P}_j\mathcal{Q}_k,\quad\quad\quad \mu=0,1,2,\cdots,15.
\end{eqnarray}
 An explicit representation of the $U(4)$ generators can be found in \cite{su4}. One can obtain the generators of $U(2,2)$ by letting the block-diagonal generators remain unchanged, while multiplying the off-diagonal generators by a factor of $i$. The generators are of the following form:
\begin{eqnarray}\label{lambda2}
\begin{pmatrix}
A&iM^{\dagger}\\
iM&B
\end{pmatrix} = 
\begin{pmatrix}
A&0\\
0&B
\end{pmatrix} 
+ i \begin{pmatrix}
0&M^{\dagger}\\
M&0
\end{pmatrix},
\end{eqnarray}
such that all the entries $A$, \(B\), and $M$ are $2\times 2$ matrices, and $A,B$ are hermitian themselves. Explicitly, we get $16$ conserved quantities which can be grouped as
\begin{equation}
\mathcal{E} \sim \begin{pmatrix}
A&0\\
0&B
\end{pmatrix}_{jk} \mathcal{P}_j \mathcal{Q}_k,
\hspace{8mm} 
\mathcal{F} \sim \begin{pmatrix}
0&M^{\dagger}\\
M&0
\end{pmatrix}_{jk} \mathcal{P}_j \mathcal{Q}_k,
\end{equation} where \(j,k = 0,1,2,3\). In Eq. (\ref{C2D}), we have collectively denoted all of them as \(\{C^\mu\}\). For example, we have
\begin{eqnarray}
\mathcal{C}^6&=&\left(\xi_1 P_2\theta_1-\xi^*_1\pi_1 X_2\right)=\left(\xi_1 \mathcal{G}_{21}-\xi^*_1\overline{\mathcal{G}}_{21}\right), \\
\mathcal{C}^7&=&-i\left(\xi_1 P_2\theta_1+\xi^*_1\pi_1 X_2\right)=-i\left(\xi_1 \mathcal{G}_{21}+\xi^*_1\overline{\mathcal{G}}_{21}\right),   \nonumber
\end{eqnarray}
where we have defined
\begin{eqnarray}
\mathcal{G}_{ij}=P_i\theta_j,\quad \overline{\mathcal{G}}_{ij}=\pi_j X_i,\quad\quad i,j=1,2.
\end{eqnarray}

However, all the conserved quantities are not independent of each other. For instance, one easily finds that \(\{\mathcal{C}^5,\mathcal{C}^7\}\sim \mathcal{C}^1\), so on and so forth.

\subsection{Classical supercharges}
Let us define the supercharges as
\begin{eqnarray}
Q_1=(P_1+P_2)(\theta_1 +\theta_2),\quad \overline{Q}_1=(\pi_1+\pi_2)(X_1 +X_2), \\
Q_2=(P_1-P_2)(\theta_1 -\theta_2),\quad \overline{Q}_2=(\pi_1-\pi_2)(X_1 -X_2), \nonumber 
\end{eqnarray} where \( \overline{Q}_1=Q_1^*\) and \(\overline{Q}_2=Q_2^*\). Clearly, the quantities (\(Q_1\), \(Q_2\)) and their complex conjugates are fermionic quantities, i.e., they are odd, because they involve multiplication of odd variables with even ones. It is easily verified that for \(Q = \frac{Q_1 + Q_2}{2}\) and \(\overline{Q} = \frac{\overline{Q}_1 + \overline{Q}_2}{2}\), one has 
\begin{eqnarray}
&&\{X_j,Q\}\sim\theta_j, \quad \{P_j,Q\} = 0, \quad \{\theta_j,Q\} = 0, \quad \{\pi_j,Q\} \sim P_j,  \\
&&\{X_j,\overline{Q}\} = 0, \hspace{4.8mm} \{P_j,\overline{Q}\} \sim \pi_j,  \hspace{1.9mm} \{\theta_j,\overline{Q}\} \sim X_j,  \hspace{0.9mm} \{\pi_j,\overline{Q}\} = 0.  \nonumber  
\label{susyt1}
\end{eqnarray}
One could similarly consider another linearly-independent combination of \(Q_1\) and \(Q_2\): \(Q' = \frac{Q_1 - Q_2}{2}\) and \(\overline{Q'} = \frac{\overline{Q}_1 - \overline{Q}_2}{2}\), which gives (index \(k\) is being summed over)
\begin{eqnarray}
&&\{X_j,Q'\}\sim (1 - \delta_{jk})\theta_k, \quad\quad \{P_j,Q'\} = 0, \label{susyt2} \\
 &&\{\pi_j,Q'\} \sim (1 - \delta_{jk}) P_k, \quad\quad\hspace{0.3mm} \{\theta_j,Q'\} = 0, \nonumber \\
&&\{P_j,\overline{Q'}\} \sim  (1 - \delta_{jk}) \pi_k, \quad\quad \hspace{0.3mm} \{X_j,\overline{Q'}\} = 0,   \nonumber \\
 &&\{\theta_j,\overline{Q'}\} \sim (1 - \delta_{jk}) X_k,  \quad\quad \{\pi_j,\overline{Q'}\} = 0.  \nonumber 
\end{eqnarray}
Clearly, sets of relations such as Eqs. (\ref{susyt1}) and (\ref{susyt2}) define SUSY transformations between the bosonic and fermionic variables of the system and can be viewed as appropriate generalizations of proposition-(\ref{prop1}). We brand these $Q$'s as the classical supercharges and have verified that they satisfy 
\begin{eqnarray}
\{Q_i,Q_j\}=0=\{\overline{Q}_i,\overline{Q}_j\}, \label{supercharge algebra 1} \\
\{Q_i,\overline{Q}_j\}=2\left(\sigma^m \mathbf{H}_m\right)_{ij},\label{supercharge algebra 2}
\end{eqnarray}
with $\sigma^m=(\sigma^0\;\sigma^3)$ and $\mathbf{H}=\left(\mathcal{P}^T(\lambda^{0})\mathcal{Q},\;\mathcal{P}^T(\lambda^{1}+\lambda^
{13})\mathcal{Q}\right)$. Eqs. (\ref{supercharge algebra 1}) and (\ref{supercharge algebra 2}) are the generalizations of Eqs. (\ref{2}) and (\ref{QQH}), respectively. \\

 It is not hard to verify that the integrals of motion from the bosonic and fermionic parts can be mapped to each other, say, using Eq. (\ref{susyt1}). For instance, if we consider 
\begin{eqnarray}
    \pi_1\theta_2=\frac{i{\xi}^*_2\xi_1}{2}\left(\mathcal{C}^{13}+i\mathcal{C}^{14}\right),\quad \quad  \pi_2\theta_1=\frac{i{\xi}^*_1\xi_2}{2}\left(\mathcal{C}^{13}-i\mathcal{C}^{14}\right),\nonumber \\
\end{eqnarray}
we obtain
\begin{eqnarray}
    \{\{\mathcal{C}^1,Q\},\overline {Q}\}=i(\pi_1\theta_2+\pi_2\theta_1),\quad \quad   \{\{\mathcal{C}^2,Q\},\overline{ Q}\}=(\pi_1\theta_2-\pi_2\theta_1).
\end{eqnarray}

Finally, we point out that the supercharges \(Q_1\) and \(Q_2\) admit an \(R\)-symmetry. In order to make this explicit, we consider a modified set of supercharges as 
\begin{eqnarray}
\begin{pmatrix}
Q_1\\
Q_2
\end{pmatrix}
\rightarrow
\begin{pmatrix}
e^{i\phi}&0\\
0&e^{i\psi}
\end{pmatrix}
\begin{pmatrix}
Q_1\\
Q_2
\end{pmatrix},
\end{eqnarray} where \(\phi, \psi \in \mathbb{R}\). The new supercharges still satisfy Eqs. (\ref{supercharge algebra 1}) and (\ref{supercharge algebra 2}) which means they admit an \(R\)-symmetry under global phase transformations from the group \(U(1) \oplus U(1)\). 

\section{Other models}\label{sec5}
In this section, we comment on the (classical) supersymmetric generalizations of two more interesting oscillator models. We first discuss the Pais-Uhlenbeck oscillator and then take up the isotonic oscillator.

\subsection{Pais-Uhlenbeck oscillator}
The Pais-Uhlenbeck oscillator is a toy model for certain higher-derivative gravity theories \cite{PU1}. 
Its equation of motion is of the fourth order and reads \cite{smilga}
\begin{equation}
\ddddot{z} + A \ddot{z} + B z = 0 ,
\end{equation} for \(z = z(t)\), where the overhead dots stand for the time-derivatives. This can be factorized in the manner,
\begin{equation}
\bigg(\frac{d^2}{dt^2} + \omega_1^2\bigg)\bigg(\frac{d^2}{dt^2} + \omega_2^2\bigg) z = 0,
\end{equation} where \(A = \omega_1^2 + \omega_2^2\) and \(B = \omega_1^2 \omega_2^2\). As is well known, 
the Pais-Uhlenbeck oscillator can be written as a system of two oscillators \cite{PU2, PU3, PU4, smilga}. While its supersymmetrization has been discussed earlier in \cite{PUS1,PUS2,PUS3}, we present a supersymmetrization based on a special `positive-energy' case for which the Lagrangian and Hamiltonian has been presented in \cite{PU3} and which resembles a system with two harmonic oscillators with positive kinetic energies. The Hamiltonian reads (cf. Eq. (12) of \cite{PU3})
\begin{equation}
H = \frac{1}{2} (p_x^2 + p_y^2) + \frac{1}{2} (\mu_1 x^2 + \mu_2 y^2 - 2 \rho xy),
\end{equation} where \(\mu_1\), \(\mu_2\), and \(\rho\) are some real and positive constants. Defining rotated coordinates and momenta according to
\begin{eqnarray}
q_1 &=& x \cos \alpha + y \sin \alpha, \label{rot} \\
q_2 &=& - x \sin \alpha + y \cos \alpha, \nonumber \\
p_1 &=& p_x \cos \alpha + p_y \sin \alpha, \nonumber \\
p_2 &=& - p_x \sin \alpha + p_y \cos \alpha, \nonumber
\end{eqnarray} we can at once diagonalize the Hamiltonian in the form
\begin{equation}\label{PUH}
H = \frac{1}{2} (p_1^2 + p_2^2) + \frac{1}{2} (a^2q_1^2 + b^2q_2^2).
\end{equation} The constants \(a^2\), \(b^2\), and \(\alpha\) can be straightforwardly related to the parameters \(\mu_1\), \(\mu_2\), and \(\rho\) \cite{PU3}. Notice that the transformations given in Eq. (\ref{rot}) stand for the \(SO(2)\) transformations acting individually on the \(x-y\) and \(p_x-p_y\) planes. It is needless to mention that the transformations are canonical. \\

We further rescale the variables as \(q_1 \rightarrow q_1/\sqrt{a}\),  \(q_2 \rightarrow q_2/\sqrt{b}\),  \(p_1 \rightarrow \sqrt{a} p_1\), and  \(p_2 \rightarrow \sqrt{b} p_2\), so that the Hamiltonian acquires a representation that reads
\begin{equation}\label{PUH1}
H = \frac{a}{2} (p_1^2 + q_1^2) + \frac{b}{2} (p_2^2 + q_2^2).
\end{equation}
Thereafter, following Eq. (\ref{can_pq}), we perform another set of canonical transformations, namely,
\begin{eqnarray}
P_1&=&\frac{p_1-iq_1}{\sqrt{2}},\quad\quad\quad\quad X_1=\frac{q_1-ip_1}{\sqrt{2}}, \\
P_2&=&\frac{p_2-iq_2}{\sqrt{2}},\quad\quad\quad\quad X_2=\frac{q_2-ip_2}{\sqrt{2}}, \nonumber
\end{eqnarray} under which Eq. (\ref{PUH1}) becomes \(H = i (a P_1 X_1 + b P_2 X_2)\). We label this as \(H_{\rm B}\), i.e., \(H_{\rm B} = i (a P_1 X_1 + b P_2 X_2)\). We are now in a position to suggest supersymmetric generalizations of the Pais-Uhlenbeck oscillator. We describe two possibilities as follows: \\

(a) Let us introduce two fermionic coordinate-momentum pairs \((\theta_1,\pi_1)\) and \((\theta_2,\pi_2)\), with a fermionic Hamiltonian \(H_{\rm F} = i (a \pi_1 \theta_1 + b \pi_2 \theta_2)\). Then, the total Hamiltonian becomes
\begin{equation}
H = H_{\rm B} + H_{\rm F} = a i \mathcal{P}_1^T \mathcal{Q}_1 + b i \mathcal{P}_2^T \mathcal{Q}_2,
\end{equation}
where $\mathcal{P}_1=(P_1 \; \pi_1)^T$, $\mathcal{P}_2 =(P_2 \; \pi_2)^T$, $\mathcal{Q}_1=(X_1 \; \theta_1)^T$, and $\mathcal{Q}_2=(X_2 \; \theta_2)^T$. It is therefore clear that the total Hamiltonian is merely two copies of Eq. (\ref{H1d}), and thereafter, the symmetry group is \(U(1,1) \oplus U(1,1)\). The classical supercharges and supercharge operators can be defined in an identical way as discussed in Sec. (\ref{sec3}). \\

(b) An alternate procedure would be to define some new variables (locally) as
\begin{eqnarray}
\tilde{X}_1&=&\sqrt{a}X_1^{\frac{1}{2}\left(1+\frac{1}{a}\right)}P_1^{\frac{1}{2}\left(1-\frac{1}{a}\right)},  \quad \tilde{P}_1=\sqrt{a}X_1^{\frac{1}{2}\left(1-\frac{1}{a}\right)}P_1^{\frac{1}{2}\left(1+\frac{1}{a}\right)}, \label{tilde1}\\
\tilde{X}_2&=&\sqrt{b}X_2^{\frac{1}{2}\left(1+\frac{1}{b}\right)}P_2^{\frac{1}{2}\left(1-\frac{1}{b}\right)},  \quad \tilde{P}_2=\sqrt{b}X_2^{\frac{1}{2}\left(1-\frac{1}{b}\right)}P_2^{\frac{1}{2}\left(1+\frac{1}{b}\right)}, \nonumber 
\end{eqnarray}
which obey the relations \(\tilde{P}_1 \tilde{X}_1 = a P_1 X_1\) and \(\tilde{P}_2 \tilde{X}_2 = b P_2 X_2\). One can straightforwardly verify that \(\{\tilde{X}_1,\tilde{P}_1 \}_{\rm P.B.} = \{\tilde{X}_2,\tilde{P}_2 \}_{\rm P.B.}= 1\), where \(\{\cdot,\cdot\}_{\rm P.B.}\) is a Poisson bracket evaluated in the \(X_1, X_2, P_1, P_2\) basis. It therefore follows that Eq. (\ref{tilde1}) is a set of canonical transformations. In terms of these new variables, we have \(H_{\rm B} = i (\tilde{P}_1 \tilde{X}_1 + \tilde{P}_2 \tilde{X}_2) \). This is rather remarkable, because we have mapped the two-dimensional (bosonic) anisotropic oscillator to a corresponding isotropic oscillator \cite{aniso} (locally). Following the presentation of Sec. (\ref{sec4}), we can now introduce fermionic variables to obtain a supersymmetric generalization of the Pais-Uhlenbeck oscillator.

\subsection{Isotonic oscillator}
The isotonic oscillator is described by a one-dimensional potential that reads \cite{ISO1,ISO2,ISO3}
\begin{equation}\label{isopot}
V(z) = a z^2 + \frac{b}{z^2} , \hspace{5mm} z>0,
\end{equation} where \(a\) and \(b\) are positive constants. The potential for \(a = 0\) is the one leading to a quantum Hamiltonian that retains its form (up to a factor) under scale transformations, \(z \rightarrow \lambda z\), where \(\lambda\) is a scale factor; this is of great interest in conformal quantum mechanics \cite{CQM}. Interestingly, this is one of the only two rational potentials in one dimension which lead to isochronous orbits (globally), the other one being the harmonic oscillator (\(b = 0\)) \cite{CV,ISO4}. Here, isochronicity refers to the lack of sensitivity of the time period of oscillation to the amplitude/energy of the system. Moreover, it  has been demonstrated that in quantum mechanics, the isotonic oscillator shares with the familiar harmonic oscillator, the feature of admitting an equispaced spectrum \cite{ISO1,ISO5}. The isotonic oscillator has also found applications in the theory of coherent states \cite{ISO6}. \\

A (classical) supersymmetric generalization of the isotonic oscillator can be achieved in the following manner; in two spatial dimensions, denoting the position and momentum vectors as \(\mathbf{r}\) and \(\mathbf{p}\), the Hamiltonian of a unit-mass particle moving in a harmonic potential reads (see also \cite{ISO7})
\begin{eqnarray}
H &=& \frac{\mathbf{p}^2 + \omega^2 \mathbf{r}^2}{2}  \nonumber \\
&=& \frac{p_r^2}{2} + \frac{\omega^2 r^2}{2} + \frac{l^2}{2r^2},
\end{eqnarray} where \(k = \omega^2\), \(r^2 = |\mathbf{r}|^2\), and \(p_r\) and \(l\) are the radial and angular momenta. In the second equality above, we have used the polar coordinates. Thus, the isotonic oscillator arises from a two-dimensional harmonic oscillator when expressed in polar coordinates. We may perform the (canonical) rescaling, \(\mathbf{r} \rightarrow \mathbf{r}/\sqrt{\omega}\) and \(\mathbf{p} \rightarrow \sqrt{\omega} \mathbf{p}\), and then define
\begin{equation}
\mathbf{P}=\frac{\mathbf{p}-i \mathbf{r}}{\sqrt{2}},\quad\quad\quad\quad \mathbf{X}=\frac{ \mathbf{r} -i \mathbf{p}}{\sqrt{2}},
\end{equation} such that \(H = i \omega \mathbf{P} \cdot \mathbf{X}= H_{\rm B}\). Extending the classical phase space to include the fermionic pairs, \((\theta_1,\pi_1)\) and \((\theta_2,\pi_2)\), and defining the fermionic Hamiltonian as \(H_{\rm F} = i  \omega (\pi_1 \theta_1 + \pi_2 \theta_2)\), the total Hamiltonian is rendered in the supersymmetric form, as in \(H = H_{\rm F} + H_{\rm B}\). Subsequently, the analysis of Section (\ref{sec4}) readily goes through.

\section{Concluding remarks}\label{discuss}
 In the present work, we have analyzed the symmetries of supersymmetric oscillators in one and two (spatial) dimensions in the framework of pseudo-classical mechanics or simply, pseudomechanics. As it turns out, they are respectively associated with symmetry groups, \(U(1,1)\) and \(U(2,2)\). It may be pointed out that in two spatial dimensions, one can in principle, have an additional presence of `anyons' which are particles that obey `any' statistics \cite{anyons}. This has to do with the fact that $\Pi_1\frac{\left({\mathbb R}^2-{\rm origin}\right)}{{\mathbb Z}_2}=\mathbb{Z}$, where $\Pi_1$ denotes the first homotopy group, ${\mathbb Z}$ is the set of integers, ${\mathbb Z}_2$ is the multiplicative group of $\pm 1$, and ${\mathbb R}^2$ is the configuration space of a single particle in two spatial dimensions. It would be interesting to see whether the description of such exotic particles fits into our formalism. However, we have not pursued such a possibility here.\\

Finally, let us remark on the possibility of defining Nambu brackets \cite{Nambu1,Nambu2,Nambu3} on the phase space containing both bosonic and fermionic variables. Consider the one-dimensional supersymmetric oscillator. Following \cite{Nambu3}, we may define a generalized Nambu bracket as
\begin{eqnarray}
\{\cdot,F_0,F_3,F_1\}=-\frac{1}{2F_2}\frac{\partial\left(\cdot,F_0,F_3,F_1\right)}{\partial\left(P,X,\pi,\theta\right)},
\end{eqnarray}
where we have
\begin{eqnarray}
&&F_0=\mathcal{Z}^0, \hspace{4mm} F_1=\frac{{\xi}^*}{2}\left(\mathcal{Z}^1+i\mathcal{Z}^2\right)+\frac{{\xi}}{2}\left(\mathcal{Z}^1-i\mathcal{Z}^2\right), \\
   && F_2 =i\left[\frac{{\xi}}{2}\left(\mathcal{Z}^1-i\mathcal{Z}^2\right)-\frac{{\xi}^*}{2}\left(\mathcal{Z}^1+i\mathcal{Z}^2\right)\right], \hspace{4mm}   F_3=\mathcal{Z}^3. \nonumber 
\end{eqnarray}

 One should notice that although the right-hand side involves a `Jacobian' which intrinsically appears to be antisymmetric, following the discussion presented in Sec. (\ref{sec2}), we must be careful to perform the left (right) derivatives with respect to the momenta (coordinates). If this prescription is followed, one can check that we get the equation of motion which reads
\begin{eqnarray}
\dot{P}=\{P,F_0,F_3,F_1\}=-i P,
\end{eqnarray} which exactly matches with the Hamilton's equation obtained using Eq. (\ref{XH}), involving the generalized Poisson bracket. One further gets \(\dot{X} = \{X,F_0,F_3,F_1\}\), \(\dot{\theta} = \{\theta,F_0,F_3,F_1\}\), and \(\dot{\pi} = \{\pi,F_0,F_3,F_1\}\), meaning that the generalized Nambu bracket describes the classical dynamics of the supersymmetric oscillator. Thus, the conserved quantities serve as Nambu-Hamiltonians in the present problem. \\

We remind the reader that in standard (non-supersymmetric) classical mechanics, the definition of the Nambu bracket is not unique \cite{Nambu3}. The same is true even in the pseudomechanics formalism, where one can pick any three amongst the conserved quantities \(\{\mathcal{Z}^i\}\) or their (suitable) linear combinations to serve as Nambu-Hamiltonians, and choose an appropriate normalization of the bracket to describe the equations of motion. A straightforward generalization to other examples is possible. We conclude this paper by noting that although we have introduced a generalized Nambu bracket for pseudomechanics, its quantum counterpart remains unclear.\\



\textbf{Acknowledgements:} 
A.S. would like to acknowledge the financial support from IIT Bhubaneswar in the form of an Institute Research Fellowship. The work of A.G. is supported by the Ministry of Education (MoE), Government of India, in the form of a Prime Minister's Research Fellowship (ID: 1200454). B.B. thanks Brainware University for infrastructural support. We thank Abhishek Chowdhury for his interest in this work.


\end{document}